\documentclass[10pt,letterpaper]{iopart}

\usepackage{amsopn}
\usepackage{amsfonts}
\usepackage{amssymb}
\usepackage{mathrsfs}
\usepackage{bbm}
\usepackage{braket}

\DeclareMathOperator{\SU}{SU}
\DeclareMathOperator{\U}{U}

\usepackage{hyperref}

\begin{document}

\renewcommand{\thefootnote}{\arabic{footnote}}	

\title{Entanglement entropy and nonabelian gauge symmetry}
\author{William Donnelly}
\address{
Department of Applied Mathematics, University of Waterloo, \\
200 University Ave. W,\\
Waterloo, Ontario, N2L 3G1, Canada
}
\ead{wdonnelly@math.uwaterloo.ca}

\begin{abstract}
Entanglement entropy has proven to be an extremely useful concept in quantum field theory.
Gauge theories are of particular interest, but for these systems the entanglement entropy is not clearly defined because the physical Hilbert space does not factor as a tensor product according to regions of space.
Here we review a definition of entanglement entropy that applies to abelian and nonabelian lattice gauge theories.
This entanglement entropy is obtained by embedding the physical Hilbert space into a product of Hilbert spaces associated to regions with boundary.
The latter Hilbert spaces include degrees of freedom on the entangling surface that transform like surface charges under the gauge symmetry.
These degrees of freedom are shown to contribute to the entanglement entropy, and the form of this contribution is determined by the gauge symmetry.
We test our definition using the example of two-dimensional Yang-Mills theory, and find that it agrees with the thermal entropy in de Sitter space, and with the results of the Euclidean replica trick.
We discuss the possible implications of this result for more complicated gauge theories, including quantum gravity.
\end{abstract}


\submitto{\CQG}
\maketitle
\ioptwocol

In quantum mechanics a subsystem can have more entropy than a larger system into which it is embedded, due to entanglement.
In \cite{Sorkin1983,Bombelli1986,Srednicki1993} it was shown that the vacuum is a highly entangled state, and it was argued that the resulting entanglement entropy could provide an explanation for the Bekenstein-Hawking entropy as arising from vacuum correlations across the black hole horizon.
In addition to its use in black hole thermodynamics \cite{Solodukhin2011}, entanglement entropy has proven to be a useful quantity in 
the AdS/CFT correspondence \cite{Ryu2006a,Ryu2006b} and in the study of topological phases of matter \cite{Kitaev2005,Levin2006}.

Entanglement entropy in a quantum field theory is typically defined as follows.
Let $\Sigma$ be a Cauchy surface partitioned into two disjoint regions $A$ and $B$ so that $\Sigma = A \cup B$.
Let us suppose that there is a factorization of the Hilbert space into a tensor product according to regions of space:
\begin{equation} \label{factorization}
\mathcal{H} = \mathcal{H}_A \otimes \mathcal{H}_B.
\end{equation}
$\mathcal{H}$ is the full Hilbert space of the theory; $\mathcal{H}_A$ and $\mathcal{H}_B$ are Hilbert spaces describing the degrees of freedom in $A$ and $B$ respectively.
Given a state $\ket\psi \in \mathcal{H}$, one can then define a reduced density matrix $\rho_A = \tr_{B} \ket{\psi}\!\bra{\psi}$ that describes physics in the causal domain of dependence of region $A$.
The entanglement entropy is the von Neumann entropy of the reduced density matrix
\begin{equation} \label{Sent}
S = -\tr(\rho_A \log \rho_A).
\end{equation}

In a theory with gauge symmetry, there is no factorization of the Hilbert space according to regions of space as in \eref{factorization}, so a fundamentally new definition of entanglement entropy is needed.
We will come to the reason for the non-factorization shortly; for now we point out that many of the systems whose entanglement entropy we are most interested in are gauge theories.
This emphasizes the importance of having a definition of entanglement entropy that would apply to theories with gauge symmetry.

One of the main uses of entanglement entropy is via the Ryu-Takayanagi formula  \cite{Ryu2006a}.
This formula is a conjectured equality between the entanglement entropy of a conformal field theory holographically dual to general relativity, and the area of certain minimal surfaces in the holographic bulk.
The best-studied version of this correspondence applies to gauge theory, specifically strongly coupled maximally supersymmetric $\SU(N)$ Yang-Mills theory.
Arguments for the conjecture make use of the Euclidean replica trick formula for the entanglement entropy \cite{Ryu2006b,Fursaev2006,Headrick2010,Hartman2013,Faulkner2013,Lewkowycz2013}, and so are only indirectly related to the canonical formula \eref{Sent}.
Given the absence of a factorization of the Hilbert space, we would like to know what canonical quantity the Ryu-Takayanagi formula is actually calculating.

In condensed matter theory, the entanglement entropy has been shown to contain useful universal information characterizing topological phases of matter \cite{Kitaev2005,Levin2006}.
In this setting one typically deals with lattice spin systems, where the factorization of the Hilbert space into regions of space is precise and unambiguous.
However a large class of these topological phases contain emergent excitations that have a low-energy description as a gauge theory \cite{Levin2005}.
As the entanglement entropy of the underlying spin system contains universal information about the emergent gauge excitations, this leads us to seek a formulation of the entanglement entropy that applies directly within the gauge theory, independent of any particular realization as a lattice model.

Perhaps the most significant role of entanglement entropy is in black hole thermodynamics \cite{Solodukhin2011}.
The original motivation for considering the entanglement entropy is that it may lead to a microscopic explanation for the Bekenstein-Hawking entropy, but so far all calculations making use of quantum field theory on a fixed classical background yield an area term that is ultraviolet divergent. 
This is not surprising; in these calculations the coupling to gravity is neglected, so one is essentially setting $G = 0$, and agreement with the Bekenstein-Hawking formula $S = A/4G$ (in $c = \hbar = 1$ units) would lead us to expect the answer $S = \infty$.
In order to obtain agreement with the Bekenstein-Hawking entropy we must introduce a nonzero $G$ in some way, and it has been argued that the gravitational interaction between quantum vacuum fluctuations will cut off the ultraviolet divergence in the entropy leading to a finite result of order $A/G$ \cite{Jacobson2012a,Bianchi2012}.
However one must then consider entanglement entropy for a quantum theory with dynamical gravity. 
Gravity is invariant under a gauge symmetry: diffeomorphism symmetry.
As in Yang-Mills theory, we expect the quantum gravity Hilbert space will not admit a factorization as in \eref{factorization}.
So one is inevitably led back to the question of what one means by entanglement entropy in a gauge theory.

The importance of entanglement entropy for gravitating systems is emphasized by the recent firewall paradox \cite{Almheiri2012}, in which the famous information loss paradox was sharpened and reformulated in terms of entanglement entropy of localized subsystems.
In this work properties of the entanglement entropy in the absence of gravity were used; one should therefore ask how the gravitational force and its associated constraints will change this argument.
In fact, the question has been posed whether it is even possible to associate entropy to a localized system in the presence of diffeomorphism symmetry \cite{Jacobson2012a}.
In \cite{Page2013,Baez2014} it was pointed out that the gauge constraints themselves lead to correlations between spacelike separated observables, and that there would be a resulting entropy of purely kinematical origin.
However this entropy was dismissed as ``fake'' or ``illusory''.
Here we will show that note only does such a kinematically mandated entropy arise naturally, but we will show that this entropy has real physical consequences, and therefore should not be considered spurious.

Let us return to the question of non-factorization of the Hilbert space for gauge theories.
First of all, why would we expect the Hilbert space of any quantum field theory to obey the factorization property \eref{factorization}?
Roughly speaking, the factorization is the quantum analogue of the statement that initial data on $\Sigma$ is in one-to-one correspondence with pairs of initial data sets on $A$ and $B$ respectively.
For example, in a scalar field one should specify the field $\phi$ and its momentum $\pi$ on $A$ and $B$ respectively, which is equivalent to specifying $\phi$ and $\pi$ on all of $\Sigma$.

Such a factorization can be made precise with a lattice regulator: 
for a scalar field theory we introduce a harmonic oscillator at each lattice site. 
The Hilbert space associated to a region $A$ is a tensor product of Hilbert spaces at each lattice site inside $A$, so that the factorization property \eref{factorization} is indeed satisfied.
This property of factorization does not strictly speaking survive in the continuum limit; the two point function must follow a universal short-distance structure in order for the state to have finite energy.
Nevertheless, having a regulator that obeys factorization allows us to define an entanglement entropy for each value of the lattice spacing.
Universal continuum results can then be extracted by subtracting divergent parts of the regulated entanglement entropy.

In gauge theory there is another reason for non-factorization of the Hilbert space.
In a Hamiltonian formulation of gauge theory the gauge transformations are generated by constraints. 
Physical states must be gauge-invariant, so they are annihilated by the constraints.
But the constraints are differential equations that must be satisfied by the initial data; their solution involves nontrivial matching between initial data in the two regions $A$ and $B$.

For example, in electrodynamics the constraint generating gauge transformations is Gauss' law, $\nabla \cdot E = 0$.
If we specify initial data on $A$ and $B$ by giving a gauge connection and an electric field, then $\nabla \cdot E$ has a singular part proportional to the difference in the normal component of the electric field $E_\perp$ across the boundary.
Thus imposing Gauss' law at the boundary implies a matching of the normal electric field; initial data in $A$ and $B$ cannot be specified independently.

This property of the continuum theory persists on the lattice.
In lattice gauge theory, the configuration space consists of group elements representing holonomies of the gauge field along edges of the lattice, and gauge constraints are imposed at the vertices.
Since the degrees of freedom live on the edges, it seems natural to define a splitting by partitioning the edges between regions.
However these degrees of freedom are not independent due to the constraints at the vertices.

Here we will describe a splitting of the Hilbert space of lattice gauge theory by
embedding the physical Hilbert space into the Hilbert space of two regions,
\begin{equation}
\mathcal{H} \to \mathcal{H}_A \otimes \mathcal{H}_B.
\end{equation} 
The Hilbert spaces $\mathcal{H}_A$ and $\mathcal{H}_B$ contain degrees of freedom that transform nontrivially under the gauge group of the boundary.
This splitting was introduced for loop quantum gravity (which can be viewed as an $\SU(2)$ gauge theory on a dynamical lattice) in \cite{Donnelly2008}, and for discrete abelian gauge theories in \cite{Buividovich2008b}. 
In \cite{Donnelly2011} the construction was generalized to abelian and nonabelian compact gauge groups on the lattice.
Recently a similar approach has been applied to two-dimensional conformal field theories \cite{Ohmori2014}.

The definition proceeds by partitioning the vertices of the lattice into two sets, analogous to the regions $A$ and $B$.
The degrees of freedom residing on edges that cross from one region into the other are split in half, with one half assigned to each region.
We introduce new degrees of freedom at the point where the edge is split; these degrees of freedom are not gauge invariant but act like surface charges confined to the entangling surface (the common interface of the regions $A$ and $B$).
The physical Hilbert space is identified with the gauge-invariant subspace of this split Hilbert space, and the entanglement entropy of a physical state is defined to be the entanglement entropy of its image under the embedding into the split Hilbert space.

Our definition of entanglement entropy refers to additional non-gauge-invariant degrees of freedom introduced at the boundary.
One might worry that this contribution to the entropy is spurious and should not be included in the entropy.
In \cite{Casini2013} an alternative definition of entropy was proposed for abelian lattice gauge theories (see also \cite{Radicevic2014}).
Rather than working at the level of Hilbert spaces, one associates to each region a subalgebra of the gauge-invariant operators, so that the entropy refers only to the measurement statistics of gauge-invariant observables. 
There is a choice of how one associates gauge-invariant operators to regions. 
One natural choice, called the electric boundary, associates to each region an algebra consisting of all electric fields on edges within the region and all holonomies of the connection around closed curves lying entirely inside the region.
As pointed out in \cite{Casini2013}, our definition of entropy coincides with entropy associated to the electric subalgebra \emph{for abelian gauge theories}.
This is encouraging, since the electric definition is the only choice that reproduces the well-established topological entanglement entropy for the toric code \cite{Kitaev1997}.
However we will see that in the nonabelian case there is a portion of the entropy that cannot be interpreted as statistical uncertainty of gauge-invariant operators.
In the nonabelian setting our definition of entanglement entropy disagrees with the entropy of the electric subalgebra.

Given that we have two disagreeing definitions of the entanglement entropy lattice gauge theory, we should devise consistency checks to distinguish between them.
Here we focus on a simple model of gauge theory, two-dimensional Yang-Mills theory.
We will be able to compare our canonical definition of entanglement entropy with existing Euclidean approaches to the entanglement entropy.
In particular we will show that our definition of the entanglement entropy agrees with the thermal entropy in de Sitter space, and with the replica trick.
The former gives an interpretation of the extra degrees of freedom in the entanglement entropy; they act just like a degeneracy of states that would arise for charges transforming in nontrivial representations of the gauge group.

We begin in section \ref{section:2D} by considering the case of two-dimensional gauge theory, first the abelian case in \ref{section:EM2}, and then Yang-Mills theory with an arbitrary compact gauge group in \ref{section:YM2}.
In section \ref{section:dS} we compare our definition of entanglement entropy with the de Sitter thermal entropy, and in section \ref{section:replica} we compare it with the result of the replica trick; in both cases the formulae are shown to agree.
Having considered two-dimensional Yang-Mills we then consider Hamiltonian lattice gauge theory in section \ref{section:lattice}.
This follows straightforwardly from the case of 2-dimensional Yang-Mills theory, since a Hamiltonian lattice gauge theory can be viewed as a 2-dimensional Yang-Mills theory associated to each edge of the lattice, glued together at the vertices.
We conclude in section \ref{section:discussion} with a discussion of possible implications beyond two-dimensional gauge theory and some areas of future work.

\section{Two-dimensional gauge theory} \label{section:2D}

To illustrate the issues that arise in entanglement entropy of gauge fields, we first consider Yang-Mills theory in two spacetime dimensions. 
This theory provides a good test case because it has non-abelian gauge symmetry and yet is simple enough to be exactly solvable.
Though the theory has no local degrees of freedom, it does have global degrees of freedom that carry a finite amount of entanglement entropy.
Thus we avoid difficult questions of renormalization of the entanglement entropy, and identifying its universal and non-universal parts.

Entanglement entropy of two-dimensional $\SU(N)$ Yang-Mills was considered in \cite{Velytsky2008,Gromov2014} using Euclidean methods.
Its canonical interpretation was discussed briefly in \cite{Donnelly2012}.
Though the technical results of this section are not new, their interpretation exposes features of the entanglement entropy that may carry forward to more interesting gauge theories.

\subsection{Two-dimensional electrodynamics} \label{section:EM2}

Let us consider first the case of two-dimensional electrodynamics, with gauge group $\U(1)$.
We will work with the canonical quantization, in which space is a circle.

The configuration degree of freedom is a $\U(1)$ connection on the circle, which can be locally expressed as a vector potential 1-form $A$.
Its conjugate variable is the electric field $E$.
However these are not independent gauge-invariant degrees of freedom. 
$\oint A$, the integral of the connection around the spatial circle, is the only gauge-invariant configuration variable.
Similarly, Gauss' law $\nabla \cdot E = 0$ implies that $E$ is constant over the circle.
Their commutation relations are $[\oint A, E] = i$, so the gauge invariant degrees of freedom form a single canonically conjugate pair.

This describes the gauge theory with gauge group $\mathbb{R}$.
For gauge group $\U(1)$, one must also quotient by large gauge transformations; this identifies the variable $\oint A$ modulo $2 \pi / q$, where $q$ is the fundamental charge.
Because $\oint A$ is a periodic variable, $E$ is quantized as $E \in q \mathbb{Z}$.
The Hamiltonian is $H = \int dx \frac{1}{2} E^2 = \frac{L}{2} q^2 n^2$ where $L$ is the circumference of the spatial circle.
Thus electrodynamics on a circle is completely equivalent to quantum mechanics on a circle (though they are not the same circle -- the latter circle is identified with the $\U(1)$ gauge group).

The Hilbert space is $L^2(\U(1))$, the space of functions on $\U(1)$ that are 
square-integrable in the Haar measure.
On this space $A$ acts as a coordinate, and $E$ acts as a momentum $E = i \partial_A$.
It is convenient to use the discrete momentum basis for $E$, whose states we can label as $\ket{n}$, with $E \ket{n} = qn \ket{n}$.

In order to discuss entanglement, we have to associate a Hilbert space not only to the whole circle but to a sub-interval.
On an interval we have only one gauge-invariant observable, the electric field $E$.
This is because the canonically conjugate variable $A$ can only be measured globally.

We can define a Hilbert space for an interval by enlarging the algebra of observables to allow for breaking of the gauge symmetry only at the endpoints of the interval.
We then have an observable $\int A$, the integral of the connection over the interval, and it is again canonically conjugate to $E$.
Thus we can assign to an open interval the same Hilbert space as for a closed circle, the space $L^2(\U(1))$.
But now this Hilbert space carries a representation of the gauge group of the interval endpoints $\U(1) \times \U(1)$.

Now suppose we have two intervals $[a,b]$ and $[b,c]$. 
How is the Hilbert space of their union $[a,c]$ related to the Hilbert spaces for subregions?
The holonomy on the union $[a,c]$ is the product of holonomies, so the configuration spaces are related by the multiplication map $\U(1) \times \U(1) \to \U(1)$.
This induces a pullback map on the wavefunctions,
\begin{equation} \label{u1map}
L^2(\U(1)) \to L^2(\U(1)) \otimes L^2(\U(1)).
\end{equation}
Concretely, when $g = g_1 g_2$, a wavefunction $\psi(g)$ induces a wavefunction $\psi'(g_1, g_2) = \psi(g_1 g_2)$.
This map is an isometry (it preserves the $L^2$ norm), which is a consequence of the invariance of the Haar measure.

While it is useful to have this more abstract point of view, there is a much more concrete version in the electric field representation.
In the electric field basis the map \eref{u1map} has the form
\begin{equation} \label{embedding}
\ket n \mapsto \ket n \otimes \ket n.
\end{equation}
In other words, the state of constant electric field $E$ on the larger interval is identified with the state having the same value of the electric field on both subintervals.

We are now in a position to calculate the entanglement entropy for an arbitrary state $\ket\psi$.
Expressing this state in the position basis, we can use the map \eref{embedding} to express it as,
\begin{equation}
\ket \psi = \sum_n \psi(n) \ket n \to \sum_n \psi(n) \ket n \otimes \ket n.
\end{equation}
The reduced density matrix is
\begin{equation}
\rho_A = \sum_n p(n) \ket{n} \! \bra{n}
\end{equation}
where $p(n) = | \psi(n)|^2$.
The entanglement entropy is straightforwardly calculated for this state, and is given by
\begin{equation} \label{Su1}
S = -\sum_n p(n) \log p(n).
\end{equation}
Note that the state $\rho$ loses the relative phase information between different $\ket n$ states. 
This is because they transform in different representations of the boundary gauge group.
This defines a superselection rule on the Hilbert space $\mathcal{H}_A$: gauge-invariant states of the electric field can be mixed but not superposed.

This result for the entropy is independent of the number of intervals considered.
For example, suppose that the region $A$ consists of several disjoint intervals.
We can express $\ket{\psi}$ as a state in the tensor product of an arbitrary number of intervals by iterating the map \eref{embedding}:
\begin{equation}
\psi \to \sum_n \psi(n) \ket n \otimes \ket n \otimes \ket n \cdots.
\end{equation}
We can see that the entropy is independent of the number of intervals traced out.
This reflects the fact that the electric field is constant over all of space.
Having access to an additional interval therefore does not change the amount of information one can acquire about the state.

The interpretation of the result \eref{Su1} is straightforward.
There is only one observable in the region $A$, which is the electric field.
An observer in region $A$ will find electric field measurements that are completely correlated with those in region $B$, as a consequence of Gauss' law.
The entanglement entropy \eref{Su1} is precisely the classical entropy associated to the distribution of possible outcomes for the electric field measurement.
However we will see that in the nonabelian case, only part of the entropy can be given this simple interpretation in terms of correlations of gauge-invariant observables.

\subsection{Two-dimensional Yang-Mills theory} \label{section:YM2}

We now generalize the above construction to Yang-Mills theory.
The configuration variable is now a $G$-connection on a circle, where $G$ is a compact Lie group.
Gauge-invariant variables can be constructed from the path-ordered exponential of the connection around the circle
$u = \mathcal{P} \exp \left[ i \oint A \right]$, and the conjugate electric field $E$, which is valued in the Lie algebra $\mathfrak{g}$ of $G$.

Unlike in electromagnetism, these degrees of freedom are not gauge invariant when $G$ is nonabelian.
In order to define $u$, we have to choose a point $x$ at which to start and end the integration, and the holonomy $u$ transforms under a gauge transformation as $u \to g(x) u g(x)^{-1}$.
The physical Hilbert space consists only of those wave functions that are invariant under this symmetry, i.e. function $\psi$ such that $\psi(g u g^{-1}) = \psi(u)$ for all $g \in G$.
This is the Hilbert space of $L^2$ class functions of $G$.

Now let us consider the Hilbert space of an interval $[a,b]$.
As in the case of electromagnetism, we allow for states that break gauge symmetry, but only at the endpoints.
This means that the Hilbert of an interval is $L^2(G)$, represented as functionals of $u = \mathcal{P} \exp \left[i \int_a^b A \right]$.
The analog of the electric field basis is the orthonormal basis $\ket{R, i, j}$ where $R$ runs over irreducible representations of the group, and $i,j$ are indices in the range $1,\ldots, \dim R$.
These states can be defined as wavefunctions of $u \in G$,
\begin{equation}
\braket{u|R,i,j} = (\dim R)^{1/2} R^i{}_j(u)
\end{equation}
where $R^i{}_j(u)$ are the matrix elements of $u$ in the representation $R$.
The prefactor $(\dim R)^{1/2}$ is to ensure normalization in the Haar measure.
The states $\ket{R,i,j}$ are not gauge invariant
under a gauge transformations of the endpoints, it transforms as
\begin{equation}
\ket{R,i,j} \to \sum_{k,l} R^i{}_k(g(b)) R^l{}_j(g(a)^{-1}) \ket{R, k, l}.
\end{equation}
Thus once again the Hilbert space carries a representation of the gauge group of the endpoints $G \times G$.

As before, the multiplication on the group induces a map
\begin{equation}
L^2(G) \to L^2(G) \otimes L^2(G).
\end{equation}
In the representation basis it is expressed as
\begin{equation} \label{embedding2}
\ket{R,i,j} \mapsto (\dim R)^{-1/2} \sum_k \ket{R,i,k} \otimes \ket{R,k,j}
\end{equation}
which is the nonabelian generalization of \eref{embedding}.

Now suppose that the interval is a closed circle.
In that case the indices $i$ and $j$ transform under the same group element and one can form a single gauge-invariant state for each irreducible representation denoted $\ket R$:
\begin{equation} \label{R}
\ket{R} = (\dim R)^{-1/2} \sum_{i} \ket{R,i,i}.
\end{equation}
These give an orthonormal basis of the physical (gauge-invariant) Hilbert space.

Let us now consider entanglement entropy of a general gauge-invariant state.
We can expand any such state $\ket \psi$ in the $R$ basis as $\ket \psi = \sum_R \psi(R) \ket{R}$.
Now suppose we divide the circle into two intervals, and consider the entanglement of the state between the two intervals.
Then using the definition of $R$ \eref{R} and the embedding \eref{embedding2}, we find
\begin{equation}
\ket \psi \mapsto \sum_R \psi(R) (\dim R)^{-1} \sum_{i,j} \ket{R,i,j} \otimes \ket{R,j,i}.
\end{equation}
The reduced density matrix for the first interval is
\begin{equation}
\rho = \sum_R p(R) (\dim R)^{-2} \sum_{i,j} \ket{R,i,j} \! \bra{R,i,j}
\end{equation}
where $p(R) = | \psi(R) |^2$.
Note that this depends only on $p(R)$ and not on the relative phase of the different $\ket{R}$ states.
This is again a result of the superselection rule: states transforming in different representations of the gauge group cannot be in superposition.

We can calculate the entanglement entropy, which is straightforward since $\rho$ is diagonal in the $\ket{R,i,j}$ basis. 
The result is
\begin{equation}
S = \sum_R p(R) (- \log p(R) + 2 \log \dim R).
\end{equation}
This differs from the abelian result \eref{Su1} by the presence of the second term that depends on the dimension of the representations; it is absent from the abelian calculation because all irreducible representations of an abelian group are one-dimensional.
We can again understand it as a result of the action of the boundary gauge group.
Because the original state $\psi$ was gauge-invariant, the reduced density matrix $\rho_A$ must commute with the action of the gauge transformations of the endpoints.
By Schur's lemma this forces it to have, for each irreducible representation $R$, a maximally mixed state of dimension $\dim R$.
This leads to an entropy $2 \log \dim R$, where the factor of two comes from the two endpoints.

We can easily extend this result to a region $A$ consisting of $n$ disjoint intervals, by iterating the map \eref{embedding2}.
The end result is an entropy
\begin{equation} \label{YM2entropy}
S = \sum_R p(R) (- \log p(R) + 2n \log \dim R).
\end{equation}
We can see that the $\log \dim R$ term appears once for each of the $2n$ points on the boundary of $A$. 
They come because each application of \eref{embedding2} introduces a maximally mixed state of dimension $\dim R$ on the index $k$.

The interpretation of the result is different for the two terms in \eref{YM2entropy}.
The first term, $-\sum_R p(R) \log p(R)$, is due to correlations of the gauge-invariant observables just as in the abelian case.
Locally one can only measure gauge-invariant functions of the electric field.
These are all functions of $R$, so the first term in the entropy captures all the entropy associated to the gauge-invariant observables.

However the entropy also has a second term, $2n \sum_R p(R) \log \dim R$, that is not associated to uncertainty in gauge-invariant observables.
Rather it comes from the extra degrees of freedom at the $2n$ endpoints that arose from relaxing gauge invariance.
Since these degrees of freedom are for fundamental reasons not measurable, the question arises whether this term should be included in the entropy.
In the next section we will consider a setting in which these degrees of freedom at the endpoints have a real physical effect, and therefore argue for their inclusion in the entropy. 
We will argue that they can be interpreted as surface charges confined to the entangling surface.

\subsection{de Sitter entropy} \label{section:dS}

The calculation of entanglement entropy in the preceding section includes a term proportional to the number of interval endpoints, and which does not have a statistical interpretation in terms of uncertainty in measurement values of gauge-invariant observables.
To understand the presence of these terms in the entanglement entropy, we consider two dimensional de Sitter space, where the entanglement entropy may be understood as thermodynamic entropy.
We will show that in order to preserve the thermal character of the vacuum in the static patch of de Sitter space, we have to associate a certain degeneracy to each value of the observable $R$.
This degeneracy can be thought of as a number of microstates that change the statistics of the observable $R$ in the thermal ensemble. 
Compatibility between thermal expectation values and vacuum expectation values gives the number of microstates as $(\dim R)^2$.
It is the logarithm of this number of microstates that gives the extra contribution to the entanglement entropy.

Consider two-dimensional de Sitter space $dS_2$ in a closed slicing, which has the line element
\begin{equation}
ds^2 = -dt^2 + r^2 \cosh^2(t/r) d\phi^2
\end{equation}
here $t$ runs from $-\infty$ to $\infty$, $\phi$ is $2\pi$-periodic, and $r$ is the de Sitter radius.
We can Wick rotate $t = i r \theta$ to a Euclidean spacetime, which is a 2-sphere
\begin{equation}
ds^2 = r^2 (d\theta^2 + \cos^2(\theta) d \phi^2).
\end{equation}

For spacetimes possessing a real Euclidean section, we can define a Hartle-Hawking vacuum \cite{Jacobson1994}, which we can view as a state on the equator $\theta = 0$.
It is defined as a Euclidean path integral over the hemisphere,
\begin{equation}
\psi(u) = \int
D A \; e^{-S_E}.
\end{equation}
The path integral is over all connections $A$ on the hemisphere such that the holonomy around the equator is $u$, weighted with the Euclidean action 
\begin{equation}
S_E = \frac{1}{4} \int d^2x \sqrt{g} \, \tr[ F^{ab} F_{ab}].
\end{equation}
This state can be explicitly evaluated in the $R$ basis, with the result\footnote{This can be obtained from the results of \cite{Cordes1994}, using the fact that the hemisphere has area $2 \pi r^2$ and the Euler characteristic of a disk, $\chi = 1$.}
\begin{equation} \label{hartlehawking}
\psi(R) \propto (\dim R) \; e^{-\pi r^2 q^2 C_2(R)}.
\end{equation}

An inertial observer in de Sitter space has access only to a bounded region of spacetime and cannot measure the integral of the connection around the hemisphere.
However the nonabelian electric field can be measured at any point.
Although the electric field itself is not gauge invariant (it transforms in the adjoint representation), one can measure gauge-invariant functions of the electric field such as the quadratic Casimir $\tr(E^2)$.
These gauge-invariant quantities are all determined by the irreducible representation $R$, so the only local observable is $R$, and it is constant over the whole spacetime by the equations of motion.

One can straightforwardly determine the statistics for measurements of $R$ in the Hartle-Hawking vacuum from the wavefunction \eref{hartlehawking}:
\begin{equation} \label{pr}
p(R) \propto (\dim R)^2 e^{- 2 \pi r^2 q^2 C_2(R)}
\end{equation}
up to an overall constant factor that normalizes the probability distribution.

We now consider an alternative derivation of this result by viewing the Hartle-Hawking vacuum as a thermal state on the static patch of $dS_2$.
Let us consider the inertial observer following the inertial trajectory  $\phi = \pi/2$.
This observer follows the flow of the Killing vector field $\xi$, which generates a de Sitter boost:
\begin{equation} \label{killing}
\xi = r \sin(\phi) \partial_t + \cos (\phi) \tanh (t/r) \partial_\phi
\end{equation}
This vector field has a bifurcate Killing horizon, with a bifurcation surface consisting of the two points at $t = 0$, $\phi = 0, \pi$.

In the present situation of a spacetime with a real Euclidean section and a bifurcate Killing horizon, a formal argument using the Euclidean path integral says that the reduced density matrix associated with the wedge is given by
\begin{equation} \label{thermal}
\rho \propto e^{- 2 \pi K_\xi}
\end{equation}
where $K_\xi$ is the generator of the de Sitter boost along $\xi$ \cite{Jacobson1994}.
The boost generator is given by an integral over the surface $t = 0, 0 \leq \phi \leq \pi$.
\begin{equation}
K_\xi = \int_{0}^\pi d \phi \sqrt{q}\; T_{ab} \, \xi^a \, n^b = \int_0^\pi d \phi \sin (\phi) \, r^2 \, T_{tt},
\end{equation}
where $\sqrt{q}$ is the determinant of the spatial metric, and $n$ is the unit normal.

In two-dimensional Yang-Mills theory, the energy density $T_{tt}$ is proportional to the square of the electric field (there is no magnetic field), which is expressible in terms of the quadratic Casimir:
\begin{equation}
T_{tt} = \frac{1}{2} \tr(E^2) = \frac{1}{2} q^2 C_2(R).
\end{equation}
Thus we obtain the boost generator in the representation basis:
\begin{equation} \label{Kxi}
K_\xi = \int_0^\pi d \phi r^2 \sin (\phi) \frac{1}{2} q^2 C_2(E(x)) = r^2 q^2 C_2(R).
\end{equation}
Unsurprisingly, the boost generator can be expressed in terms of $R$, the only local gauge-invariant operator in the theory.

From the thermal form of the state \eref{thermal}, we can find the statistics for measurements of the representation $R$. 
They are given by the Boltzmann distribution
\begin{equation}
p(R) \propto d(R) e^{- 2 \pi r^2 q^2 C_2(R)}
\end{equation}
where $d(R)$ is the dimension of the eigenspace corresponding to a given value of $R$, which is so far undetermined.
However we can determine it\footnote{In principle this determines $d(R)$ only up to an overall constant.
Such a constant can be absorbed into the path integral measure, or by adding a bare Einstein-Hilbert term to the action.
However this change would only shift $\log Z$ and hence the entropy by an overall additive constant and not affect any physics. 
The choice $d(R) = \dim(R)^2$ has the nice property that the 1-dimensional trivial representation is nondegenerate and hence the entropy vanishes in the limit $qr \to \infty$, but this choice is not forced on us by this argument.}
from the requirement that the thermal distribution match the probability distribution \eref{pr} calculated from the vacuum wave functional, giving $d(R) = (\dim R)^2$.
This gives an alternative explanation for the appearance of the $\log \dim R$ term in the entanglement entropy: in order for the Hartle-Hawking vacuum to be a thermal state, each value of the macroscopic observable $R$ must be accompanied by $(\dim R)^2$ microstates, and the logarithm of the number of microstates contributes to the thermal entropy, which is nothing but the entanglement entropy of the vacuum.

The argument that the Hartle-Hawking vacuum is a thermal state is a purely formal one, and does not specify on which Hilbert space the thermal state is defined. 
Indeed, in the usual setting of fields on a Minkowski background, the state cannot be expressed as a density matrix on any Hilbert space, rather it is a KMS state on a von Neumann algebra\cite{Bisognano1975}.
In the case of two-dimensional Yang-Mills, we have shown that the formal argument does indeed hold as a statement about density matrices on a Hilbert space, provided the boost Hamiltonian is given by $\eref{Kxi}$, and the Hilbert space is $L^2(G)$.

This thermal result suggests an interpretation of the gauge-variant states at the endpoint of the interval as surface charges.
As an analogy, we can consider the thermodynamics of a gas of particles charged under a nonabelian gauge group. 
For a particle species transforming in a $d$-dimensional representation, there will be $d$ states that are not distinguishable by any gauge-invariant operator, but there will still be a term in the entropy proportional to $\log d$ associated with these different states.
The only difference between this case and that of an uncharged gas is that the microstates of a gas are hard to distinguish as a matter of practice; here the states are in principle indistinguishable as there is no local gauge-invariant operator that would distinguish between them.

\subsection{Replica trick} \label{section:replica}

We now consider the entanglement entropy of the Hartle-Hawking state \eref{hartlehawking} for a region $A$ consisting of $n$ disjoint intervals.
In this case the entanglement entropy does not have an interpretation as thermal entropy.
However one can still calculate the entropy from the Euclidean path integral using the replica trick \cite{Callan1994}.
We will now show that the replica trick reproduces the canonical result \eref{YM2entropy}.

To calculate the entropy via the replica trick, we consider a family of ``replicated'' spacetimes parametrized by the positive integer $\alpha$, the replica index.
The replicated manifold is obtained by taking $2 \alpha$ copies of the hemisphere on which the Hartle-Hawking state is defined, which we can label by $(i,\pm)$ for $i = 0, \ldots, \alpha-1$.
The boundary of each hemisphere consists of a circle, which we partition into two regions $A \cup B$ with $A$ and $B$ each consisting of $n$ intervals.
The replicated spacetime is obtained by gluing $(i,-)$ to $(i,+)$ along $A$ and $(i,-)$ to $(i+1,+)$ along $B$.
The partition function $Z_\alpha$ on the $\alpha$-replicated spacetime is related to the reduced density matrix $\rho_A$ by $\tr(\rho_A^\alpha) = Z_\alpha Z_1^{-\alpha}$.
The replica trick calculates the entropy of $\rho_A$ by analytic continuation of this formula to a neighbourhood of $\alpha = 1$;
\begin{equation} \label{replicatrick}
S = - \left. \frac{\partial}{\partial \alpha} \tr (\rho_A^\alpha) \right|_{\alpha = 1}
= - \left. \frac{\partial}{\partial \alpha} \frac{Z_\alpha}{Z_1^\alpha} \right|_{\alpha = 1}.
\end{equation}

Let us now consider two-dimensional Yang-Mills theory in the Hartle-Hawking state.
The partition function is given by a sum over all irreducible representations of $G$ \cite{Cordes1994}:
\begin{equation}
Z = \sum_R (\dim R)^{\chi} e^{-\frac12 V q^2 C_2(R)}.
\end{equation}
As a consequence of area-preserving diffeomorphism symmetry it is a function only of the topology of the manifold, encoded in the Euler characteristic $\chi$, and its two-dimensional volume $V$.

For the $\alpha$-replicated Euclidean de Sitter spacetime the volume and Euler characteristic are given by\footnote{The latter can be determined by induction on $\alpha$, using the inclusion-exclusion formula for the Euler characteristic $\chi(X \cup Y) = \chi(X) + \chi(Y) - \chi(X \cap Y)$.}
\begin{equation}
V = 4 \pi r^2 \alpha, \qquad \chi = 2\alpha + 2n (1 - \alpha).
\end{equation}
Thus the partition function on the replicated spacetime is
\begin{equation}
Z_\alpha = \sum_R (\dim R)^{2\alpha + 2n(1 - \alpha)} e^{-2 \pi \alpha r^2 q^2 C_2(R)}.
\end{equation}
We can express this in terms of the probability distribution of the representations $R$ which is given by \eref{pr}
\begin{equation}
p(R) = (\dim R)^2 e^{-2 \pi r^2 q^2 C_2(R)} / Z_1.
\end{equation}
With this identification, we can rewrite the partition function in terms of $p(R)$
\begin{equation}
Z_\alpha = \sum_R (\dim R)^{2n (1-\alpha)} (Z_1 p(R))^{\alpha}.
\end{equation}
Applying the replica trick formula \eref{replicatrick} to this partition function yields
\begin{equation}
S = \sum_R p(R) (-\log p(R) + 2 n \log \dim R).
\end{equation}
Thus the replica trick gives an entropy that agrees with our canonical definition of entanglement entropy \eref{YM2entropy}, including the kinematic term $\log \dim R$ for each of the $2n$ interval endpoints.

\section{Lattice gauge theory} \label{section:lattice}

We briefly discuss the extension of the above results from two-dimensional Yang-Mills theory to lattice gauge theory.
The extension is straightforward, the major difference is that lattice Yang-Mills theory has local degrees of freedom. 
We will show that the entanglement entropy is the sum of the entanglement entropy of the local degrees of freedom plus a term that takes the same form as \eref{YM2entropy} for two-dimensional Yang-Mills theory.

On a lattice, the definition of the gauge-invariant Hilbert space is similar to the case of two-dimensional Yang-Mills.
To each edge of the lattice we associate a group element $u \in G$, and these transform nontrivially under gauge transformations.
It is once again convenient to use the representation basis $\ket{R,i,j}$, where now there is an irreducible representation $R_e$ assigned to each edge of the lattice, and a representation index assigned to each endpoint.
In order to pass to the gauge-invariant Hilbert space, we have to consider gauge-invariant states.
These are obtained by attaching intertwining operators to the vertices, as described in \cite{Baez1994}. 
This intertwining operator contracts all the indices that transform under the gauge transformation at a fixed vertex in a gauge-invariant way, resulting in a gauge invariant state.
An orthonormal basis of gauge invariant states is labelled by a choice of representation for each edge and a choice of intertwiner for each vertex.
This defines the spin network basis, which is the nonabelian analog of the electric field basis in electrodynamics.

In order to split the Hilbert space, we partition the vertices of the lattices into two sets $A$ and $B$.
Each edge with one endpoint in $A$ and one endpoint in $B$ is split in the middle according to the map \eref{embedding2}.
This introduces new degrees of freedom for each boundary edge.
The Hilbert space of a region $A$ is now labelled by a representation $R$ for every edge inside $A$ including the boundary edges, a choice of intertwiners on each vertex in $A$, and labels $i = 1,\ldots, \dim R_e$ on the extra vertices introduced at the boundary.

The calculation of the entanglement entropy proceeds much as in the case of two-dimensional Yang-Mills theory \cite{Donnelly2011}.
According to the map \eref{embedding}, the degree of freedom $R_e$ for boundary edges is the same on both halves of the split edge.
Let us combine the set of representations $R_e$ for all boundary edges $e \in \partial$ into a vector $R_\partial$.
Then the different choices of $R_\partial$ will have a probability distribution $p(R_\partial)$, and the fact that $R_e$ matches on both sides leads to an entropy
\begin{equation} \label{Rentropy}
- \sum_{R_\partial} p(R_\partial) \log p(R_\partial).
\end{equation}
Similarly, the embedding \eref{embedding2} leads to a state of the labels on the boundary edges that is maximally mixed, with an entropy
\begin{equation} \label{dimentropy}
\sum_{R_\partial} p(R_\partial) \sum_{e \in \partial} \log \dim (R_e).
\end{equation}
The significant difference from two dimensions is that the state in the interior is not completely determined by the value of $R_\partial$. 
For each choice of $R_\partial$, we can label each compatible state on the interior by a choice of intertwiners $I$ and representations $R$ for the interior edges.
These degrees of freedom can also be entangled; for example a glueball in the interior of $A$ could be entangled to a glueball in the interior of $B$. 
For each set of representations $R$ on the boundary, we can define a reduced density matrix $\rho(R_\partial)$ 
The entanglement entropy associated to these reduced density matrices is
\begin{equation} \label{bulkentropy}
\sum_{R_\partial} p(R_\partial) S(\rho(R_\partial)).
\end{equation}
The full entanglement entropy is the sum of the three terms \eref{Rentropy},\eref{dimentropy}, and \eref{bulkentropy}.
Note that this construction also extends to scalar or spinor matter fields at the vertices; the quantum numbers associated with the matter degrees of freedom are included in $\rho(R_\partial)$ and the entropy contributes to the bulk entropy term \eref{bulkentropy}.

\section{Discussion} \label{section:discussion}

We have presented a definition of entanglement entropy that applies to nonabelian gauge theories on the lattice.
The essential feature of this definition is introduction of Hilbert spaces $\mathcal{H}_A$ and $\mathcal{H}_B$ with additional local degrees of freedom at the entangling surface.
These degrees of freedom are not gauge-invariant but transform nontrivially under the gauge group, like surface charges.
The entanglement entropy can be split as a sum of entropy associated with the boundary degrees of freedom, and entropy associated with bulk degrees of freedom.

While it may seem strange that the entanglement entropy could exceed the total number of gauge-invariant degrees of freedom, this is well understood in topological field theories such as Chern-Simons theory.
In such a theory the Hilbert space associated to a space without boundary is finite dimensional, but the Hilbert space of a region with boundary has an infinite number of degrees of freedom, known as edge states \cite{Balachandran1994}.
It has been argued that the entanglement entropy results from counting degrees of freedom associated to edge modes \cite{Swingle2012}.
In the topological case one can derive the universal subleading correction to the area law, the topological entanglement entropy, from the replica trick \cite{Dong2008}.
In order to make sense of this negative constant term in the entropy, there must be a positive coefficient of the area term in order for the total entanglement entropy to be positive - thus in the canonical picture there must be some local degrees of freedom contributing to the entanglement entropy, whose number grows at least linearly with the boundary area.
The present work shows that this phenomenon is not a special feature of Chern-Simons theories, but also arises for any gauge theory defined with a lattice regulator.

In order to test our definition of the entropy, we have compared it to Euclidean methods for calculating the entropy in two-dimensional Yang-Mills theory.
It was shown that the multiplicity of the eigenvalues of the gauge-invariant observables is exactly what is required by agreement of vacuum expectation values with thermal expectation values in the static patch.
The agreement between entanglement and thermal entropy is encouraging, since this is an essential ingredient in both the first and second laws of black hole mechanics.

As a second check, it was shown that our canonical definition agrees with the results of the replica trick.
This is another positive suggestion that our definition of entanglement entropy will agree with results of the replica trick more broadly.
This is essential in order to give agreement with Chern-Simons theories \cite{Dong2008} and the holographic entanglement entropy \cite{Ryu2006a}, whose derivations are based on the replica trick.

We note that the $\log \dim(R)$ term in the entanglement entropy takes the form of a sum over all boundary edges; it is extensive.
This extensive term can be ignored in certain contexts.
For example, in bounding correlation functions of local operators one considers the mutual information, which is the quantity $I(A:B) = S(A) + S(B) - S(A \cup B)$ with $A$ and $B$ disjoint regions. 
In such a case, the extensive part of the entanglement entropy would drop out because one is subtracting contributions from the boundary in the same state.
Similarly, when one considers the topological entanglement entropy $S(A \cup B) - S(A) - S(B) - S(A \cap B)$ ($A$ and $B$ are no longer assumed disjoint) any extensive terms will cancel.
If we consider only a single fixed state, then we expect that any extensive term could be absorbed into a redefinition of the classical gravitational action. 
Thus we do not expect any universal quantities depending on a single state to depend on them.

Given that the $\log \dim(R)$ term does not contribute to the mutual information or topological entanglement entropy, one might worry that the extensive contribution to the entropy is completely nonuniversal (and hence unphysical).
However one could instead consider the difference in entanglement entropy of the same region in two different states, which is relevant for the dynamics of entanglement entropy.
The ultraviolet divergences are state-independent and will cancel, but state-dependent extensive terms will contribute to the change in entanglement entropy.
In fact from the point of view of horizon thermodynamics, the extensive terms are the most important; exemplified by the Bekenstein-Hawking term, they are crucial for the validity of the laws of horizon mechanics.
In this context, the generalized second law has proven to be a very useful tool for discriminating between different possible definitions of the entropy of a horizon \cite{Ford2000,Wall2011,Sarkar2013}.
This suggests that to distinguish between different proposed definitions of the entropy we should consider processes that change the configuration of nonabelian electric charges on the entangling surface and see whether the generalized second law is satisfied.
We leave this task for future work.

We note that the definition of the entanglement entropy presented here depends not only on the definition of the theory, but on its structure as a gauge theory.
However there are gauge theories that are dual under electric-magnetic duality to a theory with a different gauge group in 3+1 dimensions \cite{Montonen1977,Seiberg1994a}, or to scalar theories in 2+1 dimensions \cite{Beasley2014}.
The simplest way out of this apparent paradox is that these dualities are not local transformations, so the entanglement of local variables in one theory maps to entanglement between nonlocal variables in the other. 
However the replica trick, together with the relation between partition functions of mutually dual theories \cite{Witten1995} suggests that there is some relation between the entanglement entropy of dual theories.

Finally, we discuss the possible extension of these results to quantum gravity.
The main feature of our approach is that the Hilbert space associated to a region with boundary contains additional degrees of freedom that carry a representation of the gauge group of the boundary.
In general relativity, these are diffeomorphisms that act in a neighbourhood of the boundary.
However, unlike the case of the Yang-Mills gauge group, the diffeomorphism group does not act ultralocally, so there is a question of exactly which diffeomorphisms are represented on the Hilbert space and which remain pure gauge.
Moreover, some diffeomorphisms will move the entangling surface.
Thus in gravity we still have the question of what are the boundary degrees of freedom, and what exactly is the symmetry group associated to the boundary whose representation theory determines the form of the boundary entanglement entropy.
This approach is currently being pursued \cite{Freidel}.

\section*{Acknowledgments}

I thank Eugenio Bianchi, Horacio Casini, Matthew Headrick, Janet Hung, Ted Jacobson, Rob Myers, Guifre Vidal, Aron Wall and Yidun Wan for questions and discussions that motivated the present work.

\section*{References}

\bibliographystyle{utphys}
\bibliography{cqg}

\end{document}